\documentclass[aps,nofootinbib,twocolumn,superscriptaddress,preprintnumbers,floatfix]{revtex4-2} 
\usepackage[nodisplayskipstretch]{setspace}
\usepackage{cancel}
\usepackage{mathtools}
\usepackage{soul}
\usepackage{physics}
\usepackage{amssymb}
\usepackage{braket}
\usepackage{amsmath}
\usepackage{epsfig}
\usepackage{hyperref}
\usepackage{breakurl}
\usepackage{bm}
\usepackage{color}
\usepackage{slashed}
\usepackage{tikz}
\usetikzlibrary{decorations.markings}
\usepackage[english]{babel}
\DeclareMathOperator{\sgn}{sgn}
\makeatletter
\newcommand{\cz}{
  \mathord{\mathpalette\vaggelis@z{z}}%
}
\addto\captionsenglish{}

\newcommand\beq{\begin{equation}}
\newcommand\eeq{\end{equation}}
\def\bea{\begin{eqnarray}}
\def\eea{\end{eqnarray}}

\DeclareRobustCommand{\SkipTocEntry}[4]{}

\newcommand{\nn}{\nonumber}

\newcommand\beal{\begin{aligned}}
\newcommand\eeal{\end{aligned}}

\usepackage{xcolor}

\newcommand{\E}{\mathrm{e}}


\definecolor{darkorange}{rgb}{1,0.549,0}

\begin{document}

\preprint{DESY-24-030\\\phantom{~}}
\title{Signatures of ultralight bosons in the orbital eccentricity of binary black holes} 
\author{Mateja Bo\v{s}kovi\'c}
\affiliation{Deutsches Elektronen-Synchrotron DESY, Notkestr. 85, 22607 Hamburg, Germany}
\author{Matthias Koschnitzke}
\affiliation{Deutsches Elektronen-Synchrotron DESY, Notkestr. 85, 22607 Hamburg, Germany}
\affiliation{II. Institut für Theoretische Physik, Universit\"{a}t Hamburg, Luruper Chaussee 149, 22761 Hamburg, Germany}
\author{Rafael A. Porto}
\affiliation{Deutsches Elektronen-Synchrotron DESY, Notkestr. 85, 22607 Hamburg, Germany}

\begin{abstract}
We show that the existence of clouds of ultralight particles surrounding black holes during their cosmological history as members of a binary system can leave a measurable imprint on the distribution of masses and orbital eccentricities observable with future gravitational-wave detectors. Notably, we find that for nonprecessing binaries with chirp masses ${\cal M} \lesssim 10\,M_\odot$, formed exclusively in isolation, larger-than-expected values of the eccentricity, i.e. $e\gtrsim 10^{-2}$ at gravitational-wave frequencies $f_{\rm GW} \simeq 10^{-2}\,$Hz, would provide tantalizing evidence for a new particle of mass between $[0.5,2.5] \times 10^{-12}\,$eV in nature. The predicted evolution of the eccentricity can also drastically affect the in-band phase evolution and peak frequency. These results constitute unique signatures of boson clouds of ultralight particles in the dynamics of binary black holes, which will be readily accessible with the Laser Interferometer Space Antenna, as well as future mid-band and Deci-hertz detectors. 
\end{abstract} 
 
\maketitle

{\bf Introduction.} The birth of gravitational-wave~(GW) science 
 \cite{KAGRA:2023pio} heralds a new era of discoveries in astrophysics, cosmology, and particle physics \cite{AlvesBatista:2021eeu}. Measuring the properties of GW signals with current and future observatories, such as the Laser Interferometer Space Antenna (LISA) \cite{LISA:2022kgy}, the Einstein Telescope (ET) \cite{Maggiore:2019uih} and Cosmic Explorer (CE) \cite{Reitze:2019iox}, as well as other Mid-band \cite{Baum:2023rwc} and Deci-hertz detectors \cite{Kawamura:2020pcg}, not~only will unravel the origins of binary black hole (BBH) mergers, it also opens the possibility to discover (very-weakly-coupled) ultralight particles that are ubiquitous in theories of the early universe~\cite{Arvanitaki:2009fg,Arvanitaki:2010sy,Marsh:2015xka,Demirtas:2018akl,Mehta:2021pwf}. Notably, the mass, spin alignment, and eccentricity are expected to be correlated with formation channels, {\it isolated} or {\it dynamical}, e.g. \cite{Belczynski:2001uc,Kowalska_2011,Breivik:2016ddj,Nishizawa:2016jji,Nishizawa:2016eza,Rodriguez:2016vmx,Rodriguez:2017pec,Rodriguez:2018pss,Lower:2018seu,Randall:2019znp,Fang:2019dnh,Romero-Shaw:2020thy,Sedda:2020wzl,2021MNRAS.507.2659G,Zevin:2021rtf,Gualandris_2022,Garg:2023lfg,Saini:2023wdk,Dhurkunde:2023qoe}; whereas boson clouds (or ``gravitational atoms"~\cite{Arvanitaki:2009fg,Arvanitaki:2010sy}), formed around black holes via superradiance instabilities~\cite{1971JETPL..14..180Z,1972JETP...35.1085Z,Press:1972zz,East:2018glu,Brito:2015oca}, can produce a large backreaction on the orbital evolution. Following analogies with atomic physics~\cite{Baumann:2018vus}, the cloud may encounter Landau-Zener (LZ) resonances~\cite{Baumann:2019ztm}, or ionization effects~\cite{Baumann:2021fkf,Baumann:2022pkl,Tomaselli:2023ysb}. The presence of a cloud then leads to large finite-size effects~\cite{Baumann:2018vus,Chia:2023tle}, floating/sinking orbits~\cite{Baumann:2019ztm}, as well as other sharp features~\cite{Baumann:2022pkl}, that  become unique signatures of ultralight particles in the BBH dynamics.\vskip 4pt For the most part, up until now backreaction effects have been studied under the simplified assumption of planar, quasi-circular orbits. The reason is twofold~\cite{Baumann:2018vus}. First, several formation scenarios lead to spins that are parallel to the orbital angular momentum \cite{Rodriguez:2016vmx}. Second, the decay of eccentricity through GW emission in vacuum~\cite{Peters:1963ux,Peters:1964zz} is expected to have circularized the orbit in the late stages of the BBH dynamics. We retain here the former but relax the latter assumption. As we shall see, adding eccentricity not only introduces a series of {\it overtones}~\cite{Berti:2019wnn,Tomaselli:2023ysb,us}, it can also have a dramatic influence in the orbital dynamics as the cloud transits a LZ-type transition. 
 Although the strength of the new resonances is proportional to the eccentricity, depending on their position and nature (floating or sinking), a small departure from circularity can lead to transitions that not only would deplete the cloud, but also induce a rapid growth of eccentricity toward a large critical (fixed-point) value: $e_{\rm cr} \in [0.3,0.6]$. As measurements of the eccentricity are correlated with formation channels, the predicted increase can impact the inferred binary's origins. Measurements of larger-than-expected eccentricities would then provide strong evidence for the existence of a new ultralight particle in nature. In particular, because of the critical fixed point, a fraction of the BBHs undergo a rapid growth of orbital eccentricity to a common value. As a result, the distribution of masses and eccentricities may feature a skewed correlation by the time they reach the detector's band.~Furthermore, for chirp masses ${\cal M} < 10 M_{\odot}$ and spin(s) aligned with the orbital angular momentum---expected to exclusively form in the field---the presence of a boson cloud at earlier times can shift a fraction of the population toward values of $e \gtrsim 10^{-2}$ at $10^{-2}$Hz, readily accessible to LISA~\cite{LISA:2022kgy}.  Furthermore, the GW-evolved eccentricity may also be within reach of the planned mid-band \cite{Baum:2023rwc} or Deci-hertz \cite{Kawamura:2020pcg} observatories. For all such events, a new ultralight boson of mass $[0.5,2.5] \times 10^{-12}$~eV forming a cloud and decaying through a LZ-type transition prior to detection, may be the ultimate culprit.\vskip 4pt 
 
The more drastic evidence is given when the resonant transition occurs in band with measurable frequency evolution. A plethora of phenomena are discussed in \cite{Baumann:2018vus,Baumann:2019ztm} for circular orbits. In addition to overtones, the increase in eccentricity would imply that higher harmonics become more relevant, which in turn affects the peak frequency of the GWs, even for floating orbits. We point out here some salient features and elaborate further on the details elsewhere~\cite{us}. \vskip 4pt

 {\bf The gravitational atom.} Ultralight particles of mass $\mu$ can form a cloud around a rotating black hole of mass $M$, via superradiant instabilities~\cite{Arvanitaki:2009fg,Arvanitaki:2010sy}. The typical mass of the (initial) cloud scales as $M_\mathrm{c,0}/M \simeq \alpha$, whereas its typical size is $r_c \simeq \tfrac{r_g}{\alpha^2}$, with $r_g \equiv \tfrac{GM}{c^2}$, and
\begin{equation}
\alpha = \frac{GM\mu}{\hbar c} \simeq 0.1 \left(\frac{M}{15 M_{\odot}}\right)\left(\frac{\mu}{10^{-12}\rm eV}\right)\,.
\end{equation}
The (scalar) cloud evolves according to a Schr\"odinger-like equation~\cite{Detweiler:1980uk,Baumann:2019eav}, with eigenstates $\ket{a}\equiv\ket{n_a l_a m_a}$, and $(n,l,m)$ the principal, orbital and azimuthal angular momentum, “quantum numbers”. (For vector clouds~\cite{East:2017mrj,Baumann:2019ztm,Baumann:2019eav}, we must include the total angular momentum.)\vskip 4pt The energy eigenvalues of the cloud scale as $\epsilon_{n l m}=\mu\left(1-\tfrac{\alpha^2}{2n^2}+f_{n l} \, \alpha^4 +h_{n l}\, \tilde a\, m \,\alpha^5\right)$, with $\tilde a$  the dimensionless spin of the black hole, see~\cite{Baumann:2019eav}. At saturation, we have $\tilde a \simeq \alpha$, whereas the combined system black hole plus cloud may still be rapidly rotating. One of the main difference w.r.t. ordinary atoms, however, is the presence of a decay/growing time, $\Gamma^{-1}_{n l m} \propto \mu \,\alpha^{4l+5}$, for a given eigenstate~\cite{Detweiler:1980uk,Dolan:2007mj,Arvanitaki:2010sy,Baumann:2019eav}. The (scalar) cloud may be populated by the dominant {\it growing} mode, $\ket{211}$, or an {\it excited} state, $\ket{322}$. Depending on $\alpha$, they may be robust to GW emission (from the cloud itself) on astrophysical scales ~\cite{Arvanitaki:2010sy,Yoshino:2013ofa,Brito:2014wla,Arvanitaki:2014wva,Brito:2017zvb,Siemonsen:2022yyf}. They can also deplete through resonant transitions in binaries \cite{Baumann:2018vus,Baumann:2019ztm}, as we discuss here. In what follows we work with $G=\hbar=c=1$ units. 

\vskip 4pt {\bf Gravitational collider goes eccentric.} 
Following ~\cite{Baumann:2018vus,Baumann:2019ztm} we consider a boson cloud around a black hole of mass $M$ in a bound orbit with a companion object of mass $M_\star$, with $q\equiv M_\star/M$ the mass ratio. The coordinates are centered at the black hole plus cloud system, with $R_\star$ the radial distance to the perturber, and $\varphi_\star$ the azimuthal angle. We consider planar motion with the spin parallel to the orbital angular momentum, with the orbit described by the semi-major axis $a$ and the eccentricity $e$, while $\varphi_\star$ corresponds to the true anomaly. We take the orbital frequency to be positive such that the two, co-rotating and the counter-rotating, orientations are identified by $\dot{\varphi}_\star=s |\dot{\varphi_\star}|$, with $s=\pm 1$.\vskip 4pt 
The gravitational perturbations of the companion induce  mixing of the atomic levels. For a perturber outside of the cloud $R_\star \gg r_\mathrm{c}$ the off-diagonal matrix elements of the Hamiltonian, $\bra{a} V_\star \ket{b} $, are given by a multipole expansion that can be written as an harmonic series~\cite{Baumann:2018vus,Baumann:2019ztm}
\begin{eqnarray}\label{eq:transition_amp}
\bra{a} V_\star \ket{b}_{l_\star}= \sum_{|m_\star| \leq l_\star} \eta^{(m_\star)}_{ab} \E^{ -i m_\star \varphi_\star } \,,
\end{eqnarray}
with $ \eta^{(m_\star)}_{ab} \propto R^{-(l_\star+1)}_\star$. The matrix elements obey selection rules 
which determine possible transitions, which we refer as hyperfine (only $\Delta m \neq 0$), fine ($\Delta \ell \neq 0,\Delta n=0$), and Bohr $(\Delta n \neq 0)$, respectively~\cite{Baumann:2018vus,Baumann:2019ztm}.\vskip 4pt

For illustrative purposes, we consider a two-level model.  The Hamiltonian equation is given by 
\begin{equation} \label{eq:Schr}
	i\begin{pmatrix}
    	\dot c_a\\\dot c_b
	\end{pmatrix} = \begin{pmatrix}-\tfrac{\Delta \epsilon}{2} &&
 	\eta_0 \left(
    \tfrac{R_\star}{R_0} \right)^{-(l_\star+1)} 
    \E^{i\Delta m \varphi_\star}
    \\\\
    {\rm c.c.} 
    && \tfrac{\Delta \epsilon}{2}-i\Gamma_b
	\end{pmatrix}\begin{pmatrix}
    	c_a\\c_b
	\end{pmatrix} \,, \\
\end{equation}
with $\Delta m \equiv m_b-m_a$, $\Delta \epsilon \equiv\epsilon_b-\epsilon_a$ the energy split, $\Gamma_b$ the width of the decaying mode, and $ \eta_0$ the value of the perturbation at a reference point $R_0$. Furthermore, since (vacuum) GW emission is expected to reduce the initial eccentricity prior to encountering the resonant transition, and for the purpose of analytical understanding, in what follows we describe the orbital evolution in the  Hamiltonian, ${\cal H}$, of \eqref{eq:Schr} using a small-eccentricity approximation,\footnote{This approximation turns out to be extremely accurate, upon comparison with numerical solutions for generic (planar) orbits, for most of the parameter space. (See appendices.)}
\begin{align} \label{eq:small_e_orbital}
\varphi_\star &\simeq \vartheta+2 e \sin \vartheta \,,\,
R_\star \simeq a (1- e \cos \vartheta) \,, \\
\dot\vartheta &\equiv s \Omega\,\,,\, \Omega=\sqrt{M(1+q)/a^3}\,, 
\end{align}
in terms of $\vartheta$, the mean anomaly~\cite{Tremaine_Dynamics}, and
apply the Jacobi-Anger expansion into Bessel functions. Hence,
\begin{eqnarray} \label{eq:Schr_ecc}
	&& \mathcal{H} =  \mathcal{D} + \sum^{\infty}_{k=-\infty}\begin{pmatrix}&&
   \eta_k \E^{i (k+\Delta m) \vartheta}
	\\\\
 	\eta_k \E^{-i (k+\Delta m) \vartheta} && 
    \end{pmatrix} \,, \\
 %
 && \mathcal{D}= \begin{pmatrix} -\frac{\Delta \epsilon}{2}  &&
	\\\\
 	 && \frac{\Delta \epsilon}{2} - i \Gamma_b
    \end{pmatrix} 
 \,, \, \eta_k \sim \eta_0 
   \frak{f}^{\frac{2}{3}(l_\star+1)} \frac{e^{|k|} }{|k|!},\,\,\, \frak{f} \equiv \frac{\Omega}{\Omega_0}\,,\nonumber
\end{eqnarray}
where we traded distance for orbital frequency. The case $(e, \Gamma_b) =0$ was studied in \cite{Baumann:2019ztm}. The slow GW-induced evolution of the orbital frequency, $\Omega(t) \simeq \Omega_0 + \gamma_0 t$ with $\gamma_0= \frac{96}{5} q M^{5/3} \frac{\Omega_0^{11/3}}{(1+q)^{1/3}} $, leads to a LZ transition~\cite{landau,Zener:1932ws} between the energy levels. The transition is triggered for \beq \Omega_0 = s \frac{\Delta \epsilon}{ \Delta m} >0\,.\eeq 
This value, dictated by the spectrum of the cloud, will serve as our reference point in the evolution of the binary.\vskip 4pt Ignoring backreaction effects (see below), the LZ solution is controlled by the parameter $z_0 \equiv \eta^2_0/(\gamma_0|\Delta m|)$, which determines the adiabaticity of the transition. As famously demonstrated in~~\cite{landau,Zener:1932ws}, starting in the far past from the $\ket{a}$ state, in the limit $2 \pi z_0 \gg 1$, the  eigenstate of the (time-dependent) Hamiltonian yields a complete population transfer into the (decaying) $\ket{b}$ mode in the far future. As it turns out, although the solution changes at finite time, controlled by the parameter $v_0 \equiv \Gamma_b/\sqrt{\gamma_0 |\Delta m|}$, the asymptotic properties of the system (ignoring backreaction) are remarkably robust against the value of the decaying width for the $\ket{b}$ state~\cite{PhysRevA.46.4110}.\vskip 4pt 

For eccentric orbits, the evolution in \eqref{eq:Schr_ecc} also features a transition at $\Omega_0$ (for $k=0$). However, it introduces a series of overtones
\begin{equation} \label{eq:freq_overtone}
 \Omega_k = \frak{f}_k \,\Omega_0\,, \quad
 \frak{f}_k = \frac{\Delta m}{\Delta m + k}  \,, \quad k \in \mathbb{Z}\,.
\end{equation}
Provided each $k$-resonance is sufficiently narrow, we can ignore the other ($k^\prime \neq k$) terms in~\eqref{eq:Schr_ecc}. As in~\cite{Baumann:2019ztm}, we can linearize the orbital evolution near the transition, $\Omega(t) =\Omega_k+f(e) \gamma_k t$, where $f(e)=\frac{1+\frac{73 e^2}{24}+\frac{37 e^4}{96}}{(1 - e^2)^{7/2}}$ and $\gamma_k \equiv \gamma_0 \frak{f}^{11/3}_k$, such that the LZ solution now depends on the modified $z_k \equiv \frac{\eta_k^2}{f(e)\gamma_k|\Delta m + k|}$ and $v_k \equiv \frac{\Gamma_b}{\sqrt{f(e)\gamma_k|\Delta m + k|}}$, respectively. 
\vskip 4pt

{\bf Orbital backreaction.}  Dissipative effects, such as GW emission, from the binary \cite{Baumann:2019ztm,Brito:2023pyl} or the cloud itself \cite{Yoshino:2013ofa,Brito:2014wla,Arvanitaki:2014wva,Brito:2017zvb,Siemonsen:2022yyf}, ionization \cite{Baumann:2021fkf,Baumann:2022pkl,Tomaselli:2023ysb,Brito:2023pyl}, and decay widths ~\cite{PhysRevA.46.4110, 1997PhRvA..55.2982V,Takahashi:2023flk}, strongly influence the LZ phenomenology, and vice versa. We focus here on the prevailing case of two-body GW emission, with the companion {\it outside} of the cloud, thus focusing on (hyper)fine resonances, combined with a two-level LZ transition into a decaying mode.\vskip 4pt The orbital dynamics is governed by flux-balance equations at infinity~\cite{Peters:1964zz, Baumann:2019ztm,Takahashi:2021yhy,Takahashi:2023flk,Tomaselli:2023ysb,us}, and at the black hole's horizon~\cite{Arvanitaki:2009fg,Ficarra:2018rfu,Ficarra:2018rfu,Hui:2022sri}:
\begin{eqnarray} \label{eq:balanceE}
\dot E_\mathrm{o}+\dot E_{\mathrm{c}}+\dot M &=& {\cal F}_{\mathrm{GW}} \equiv - \frac{32f(e)}{5} \frac{M^5 q^2 (q+1)}{a^5}  \,, \\
\dot L_\mathrm{o}+s (\dot L_{\mathrm{c}}+\dot S) &=& {\cal T}_{\rm{GW}} \equiv \frac{{\cal F}_{\mathrm{GW}}}{\Omega} \frac{g(e)}{f(e)}   \,,\label{eq:balanceL} \\
\dot M = 2 \Gamma_{b} E_{\mathrm{c}(b)}&,&\,\, \dot S = 2\Gamma_{b} L_{\mathrm{c}(b)}  \,,\label{eq:balance_hor_BH} 
\end{eqnarray}
with $g(e)=\tfrac{1+\frac{7 e^2}{8}}{(1-e^2)^2}$, and $
\dot M$, $\dot S$ the change of mass and spin due to the decay of the $\ket{b}$ state onto the black hole.\footnote{We ignore here effects due to changes in the background $(M,S)$ black hole's parameters. We justify this approximation in the appendices, see also \cite{us}.} The orbital energy and angular momentum are given by $E_\mathrm{o}= - \tfrac{ M^2 q}{2a}$ and  $L^2_\mathrm{o}= \tfrac{(M^5 q^3)(1-e^2)}{2(q+1)|E_\mathrm{o}|}$, while for the cloud is a sum over the populated states, $E_{\mathrm{c}(i)}\equiv (M_\mathrm{c,0}/\mu) \epsilon_i |c_i|^2$, and similarly for $L_{\mathrm{c}(i)}$ with $\epsilon_i \to m_i$.\vskip 4pt 

The above equations can then be rewritten as
\begin{eqnarray} 
\frac{d\Omega}{dt} &=& r \gamma_0 \frak{f}^{11/3} f(e)  \label{eq:balanceW} \,,\\
r &\equiv&  \frac{\dot{E}_\mathrm{o}}{{\cal F}_{\mathrm{GW}}} = 1 - b \frac{\mathrm{sgn}(s \Delta m )\frak{f}^{-11/6}}{\sqrt{f(e)\gamma_0 |\Delta m + k|}}      \frac{d|c_a|^2}{dt}\,, \label{eq:r_of_t} \\
\label{eq:balanceecc}
\frac{de^2}{dt} &=& \frac{2}{3} \frak{f}^{8/3}  \frac{\gamma_0}{\Omega_0} f(e) \sqrt{1-e^2} \times\\
&&\left[r \left(\frak{f}-\sqrt{1-e^2}\right)-\frak{f}+\frac{g(e)}{f(e)}\right]\nonumber \,, 
\end{eqnarray}
in terms of the orbital parameters, where
\begin{equation}
b  \equiv    \frac{3M_\mathrm{c,0}}{M} \frac{|\Delta m|\frak{f}^{-3/2}}{|\Delta m + k|^{-1/2}} \frac{(1+q)^{1/3}}{ \alpha q}  \frac{\left(M \Omega_0\right)^{1/3}\Omega_0}{\sqrt{\gamma_0}} \,,
\end{equation}
parameterises the backreaction effects on the orbit due to the cloud. It is worth emphasising that the above equations apply to generic (planar) motion, regardless of the value for the eccentricity. As we shall see, even for small initial conditions, the orbit is affected by large backreaction effects due to the presence of the cloud.\vskip 4pt As anticipated by the analysis in~\cite{Baumann:2019ztm} for the case of circular orbits (which we encourage the reader to consult for further details),  ``effective" LZ parameters emerge: $\zeta_k(t) \equiv z_k/r(t)$ and $w_k(t) \equiv v_k/\sqrt{r(t)}$, making it a fully nonlinear system. We can nonetheless estimate the value of the energy-momentum transfer near the resonance by self-consistently solving the condition $\zeta_k=z_k/r_k(\zeta_k)$. For  moderate-to-large population transfer $(\zeta_k \gtrsim 1)$, we find the limiting results:
\begin{align}
r_k &\simeq \left( 1 -  \sgn(s\Delta m) \frac{b_k}{4 \sqrt{z_k}}\right)^{-1}\,, \quad\quad (w_k \ll \zeta_k)\label{rsmallv} \\
%
r_k &\simeq 2 \left(1+\sqrt{1-\sgn(s\Delta m) \frac{b_k}{z_k v_k}} \right)^{-1}
 \,,\, (w_k \gg \zeta_k) \,. \nonumber
\end{align}

As discussed in \cite{Baumann:2019ztm}, the orbital evolution branches into either floating $(r \simeq 0)$, for $s \Delta m  <0 $, or sinking orbits $(r \gtrsim 1)$, for $s\Delta m  > 0$. However, except for the trivial case when  $\zeta_k \ll 1$,  due to the nonlinear nature of the problem the transfer of energy and angular momentum from the cloud to the orbit does not simply reduce to the quest for adiabaticity of the LZ transition, not even for $w_k \ll \zeta_k$. For instance, for extreme cases, with $z_k \gg 1$, the (unperturbed) transition spreads over long time scales, $\Delta t_\mathrm{LZ} \simeq  4 \sqrt{z_k/\gamma_k}$~\cite{Vitanov:1998yn}, which in turn reduces the orbital impact, as we see in \eqref{rsmallv}. As it turns out, in the large backreaction scenario,
the sweet spot for floating orbits occurs when $b_k \gg \sqrt{z_k}$. Even though, due to the properties of the LZ solution, a strong decay width ($w_k \gg \zeta_k$) does not alter this picture, the impact on the orbit evolution as well as the population transfer becomes suppressed by $1/v_k$, as shown in \eqref{rsmallv}. On the other hand, for the sinking case,  the largest values of $r_k$ are obtained for nonadiabatic transitions.\vskip 4pt

\begin{figure*}[t!]
\begin{tabular}{cc}
\includegraphics[width=.41\textwidth]{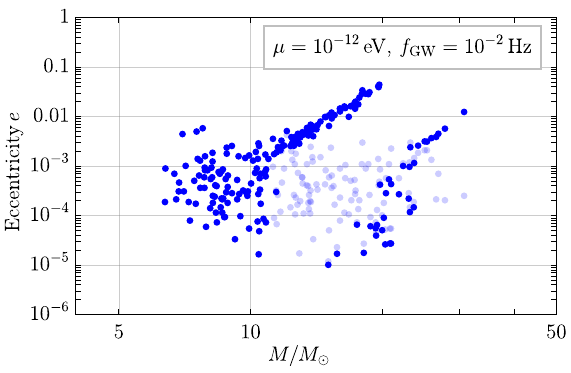}
\qquad
\includegraphics[width=.39\textwidth]{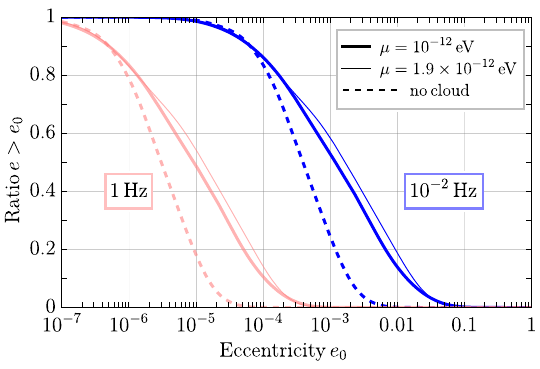}
\end{tabular}
    \caption{BBH eccentricities at $f_{\rm GW}= 10^{-2}$Hz ({\it left}), evolved with a uniformly distributed $q\in\left[0.1,1\right]$ and a boson cloud on the heavier black hole.  The pale blue dots account for the values without a cloud \cite{Breivik:2016ddj}. (BBHs with $e\lesssim 10^{-6}$ are not shown.) Cumulative effect, i.e. the ratio of binaries with eccentricities above a given value $e_0$, ({\it right}), with (solid) and without (dotted) a cloud, both at $10^{-2}\,\mathrm{Hz}$ (blue) and $1\,\mathrm{Hz}$ (pink), respectively.}
    \label{fig:Populations}
\end{figure*}

{\bf Eccentric fixed point.} For the GW-dominated epochs, with $r\simeq 1$, the leading order term in~\eqref{eq:balanceecc} vanishes, and the first contribution is at ${\cal O}(e^2)$. Likewise for the $k=0$ (main) resonance, for which the first term is $\propto \big(\tfrac{r}{2} - \tfrac{11}{3}\big)e^2$. As a result, the eccentricity is damped unless the orbit gets a large kick ($r \gtrsim 7.3$). As the influence of the cloud increases, the RHS of~\eqref{eq:balanceecc} asymptotes (modulo a positive prefactor) to $(\frak{f}_k-1)(r-1)$, in which case it enters at leading order. Moreover, the differences in the GW fluxes in~\eqref{eq:balanceE} and~\eqref{eq:balanceL} generate a distinction between the early and late resonances. In~the floating case, with $r \simeq 0$, the eccentricity grows for the early resonances $(\frak{f}_k < 1)$ and decays for the late ones $(\frak{f}_k >1)$. This can be understood by noticing that, when $\dot E_\mathrm{o}\simeq 0$, we have $\dot L_\mathrm{o} \propto \left( \tfrac{\Omega-\Omega_0}{\Omega \Omega_0} + \mathcal{O}(e^2) \right), \,$
and using $d(L^2_\mathrm{o}) \propto - d(e^2)$ the eccentricity grows for $\Omega_k < \Omega_0$ and decays whenever $\Omega_k \geq \Omega_0$. This trend is reversed in the sinking case.\vskip 4pt 

Because of the changes in the evolution of the eccentricity across different resonances, it is instructive to look at the opposite limit $e \to 1$. In that case, the RHS of~\eqref{eq:balanceecc} becomes $\propto \frac{r-1}{(1-e)^3}$. Let us consider the case of a floating orbit. Since the sign of $\frac{de}{dt}$ is positive for $\Omega_k < \Omega_0$, but turns negative when the eccentricity approaches $e \simeq 1$, this implies the existence of a critical ``attractor" fixed point, $e_{\rm cr}$, given by the condition $g(e_{\rm cr})/f(e_{cr})= \frak{f}_k$ [cf.~\eqref{eq:balanceecc}]. For~instance, 
\begin{equation}
e_{\rm cr}=\{0.46, 0.35,0.29\}, \quad {\rm for}\quad  |\Delta m|=\{1,2,3\}\,,
\end{equation}
with $k=-1$. Similarly, an unstable fixed point develops for the earlier and main sinking resonances.\vskip 4pt

For the case of floating orbits (with $s\Delta m <0$), if the backreaction is sufficiently effective to enforce $r_k \simeq 0$ while the eccentricity approaches the critical point, one can then estimate the floating time $\Delta t_\mathrm{FL} \simeq b_k/\sqrt{\gamma_k}$, left-over population $|c_a(\infty)|^2 \lesssim r_k$, and notably the growth of the eccentricity upon exiting the resonant transition, 
\begin{equation}
\label{growth-ecc}
e_{\mathrm{fin}}\simeq e_{\mathrm{cr}} \sqrt{1-\E^{-\mathrm{C_k}}} \,,\quad {\rm with} \, \,\mathrm{C_k} \sim \tfrac{\sqrt{\gamma_k}}{\Omega_k}b_k\,.
\end{equation}
Although we have used a small-eccentricity approximation to describe the initial stages of the cloud's evolution in \eqref{eq:Schr_ecc}, we have demonstrated through numerical studies that the behavior described above remains valid for generic (planar) orbits. See \cite{us} and the appendices for details.\vskip 4pt 

\begin{figure}[t!]
    \centering
    \includegraphics[width=0.35\textwidth]{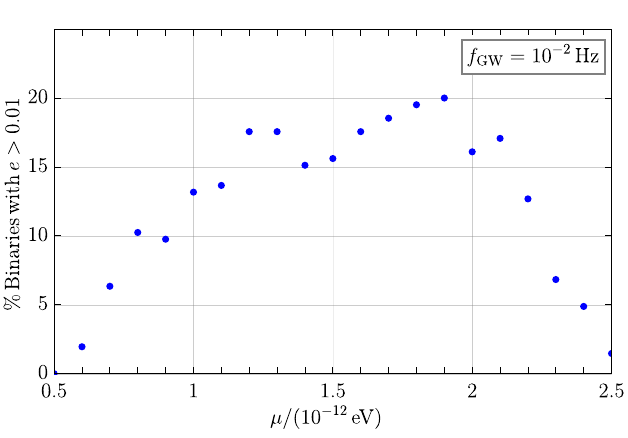}
    \caption{Percentage of binaries with eccentricities above 0.01 at $f_\mathrm{GW}=10^{-2}\,$Hz for different values of $\mu$.}
    \label{fig:muPlot}
\end{figure}

{\bf The cloud's eccentric fossil.} As it was argued in the literature \cite{Breivik:2016ddj,Nishizawa:2016eza,Nishizawa:2016jji}, the distribution of masses and eccentricities observed with LISA can in principle distinguish between formation channels. However, the contrast between vacuum evolution and the large eccentricities produced by the cloud's resonant transition can lead to dramatic changes in the expected evolution of the system. As a proof of concept, we take the stellar-mass BBH population studied in \cite{Breivik:2016ddj}, with chirp masses ${\cal M} \lesssim 10 M_\odot$, expected to form exclusively in isolation, and with the spins aligned with the orbital angular momentum. As~a consequence, the assumption of equatorial (uninclined) motion may be implemented without loss of generality (as done in \cite{Breivik:2016ddj}).\vskip 4pt  
We consider clouds of ultralight bosons of mass between $10^{-13}$ and $10^{-11}$~eV, surrounding black holes in co-rotating~orbits. Superradiance may then excite the $\ket{322}$ state which, depending on the parent black hole's mass and birth orbital frequency, will experience a series of (hyper)fine transitions.\footnote{For the BBH population and values of $\mu$ we consider here, the largest impact on the orbital evolution happens for $\alpha \gtrsim 0.1$. Hence, the overall effect from the earlier, but shorter-lived, $\ket{211}$ component of the cloud becomes subdominant.} To illustrate the distinct physical effects, and following \cite{Breivik:2016ddj,Kowalska_2011}, we consider a birth orbital frequency (for the cloud+BBH system) at $\Omega_{\rm ini}/\pi \simeq 10^{-4}\,$Hz,\footnote{Even in cases where the birth orbital frequency of the binary system may be lower, stellar evolution \cite{Belczynski:2017gds} can also lead to {\it younger} (secondary) black holes carrying the cloud.} and evolve, using the peak GW frequency \cite{Wen:2002km} 
\begin{equation}
f_{\rm GW} \simeq \frac{\Omega}{\pi} \frac{(1+e)^{1.1954}}{(1-e^2)^{3/2}}\,,    \label{peak}
\end{equation}
until $ f_{\rm GW} = 10^{-2}$ and 1 Hz. The final distribution is shown on the left panel of Fig.~\ref{fig:Populations}. While some of the BBHs experience an early overtone of the hyperfine transition, the majority are affected by the fine overtones instead. The BBHs then float over a period of time while increasing the orbital eccentricity. Moreover, the cloud typically either terminates there or decays later at the $k=0$ resonance. Depending on the parameters, the ultimate decay may decrease the eccentricity or have a small impact on the orbit.  As a result, a {\it wedge-type} distribution emerges, with the heavier black holes (within each wedge) subject to the largest increase in eccentricities.\vskip 4pt The cumulative effect is shown on the right panel of Fig.~\ref{fig:Populations}, where a significant fraction of the population (with different parent masses) is affected by the resonances, yielding values of the eccentricities at 1Hz that may be within reach of mid-band and Decihertz detectors. As the value of $\mu$ increases (decreases) the location of the wedge in the distribution moves toward lower (higher) masses. The dependence on the value of $\mu$ for this population of BBH (with ${\cal M}\lesssim 10M_{\odot}$), reaching $e\gtrsim  10^{-2}$ at $f_{\rm GW} = 10^{-2}\,$Hz, is shown in Fig.~\ref{fig:muPlot}.\vskip 4pt

\begin{figure}
    \centering
    \includegraphics[width=0.4\textwidth]{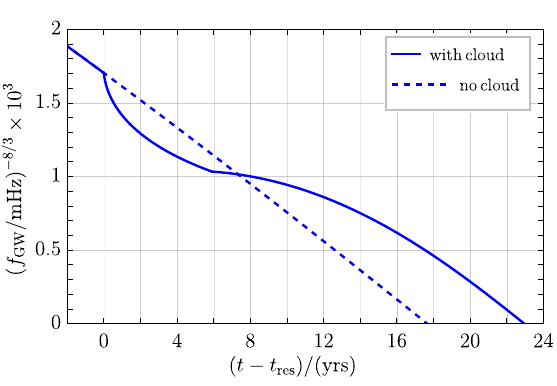}
    \caption{Evolution of the peak frequency through a resonant transition (at $t-t_\mathrm{res}=0$) in the LISA band of a GW170809-like event, compared to the evolution without the cloud.}
    \label{fig:plotInBand}
\end{figure}

{\bf Eccentric in band.} Because of the connection to formation channels, we discussed a sub-population of BBHs. However, similar conclusions apply to black holes with higher masses \cite{us}. For instance, for a GW170809-type event~\cite{LIGOScientific:2018mvr} (${\cal M} \simeq 24 M_\odot)$, with a parent black hole $M \simeq 20 M_\odot$ (carrying the cloud) and a (heavier) companion $M_\star \simeq 40 M_\odot$, we find $ f_{\rm GW} \gtrsim 10^{-2}$\,Hz at the $k=-1$ fine transition for $\mu \simeq 1.5\times 10^{-12}$ eV. The BBH reaches the resonance and floats, with approximately constant orbital frequency for about six years, while the eccentricity increases from $e\lesssim 10^{-2}$ to $e\simeq 0.1$, and likewise the peak GW frequency grows, while the cloud depletes. The resulting frequency evolution till merger, which is distinct from the growth of the eccentricity that may occur due to other astrophysical mechanisms~\cite{Randall:2018qna,Randall:2019sab},\footnote{For instance, compare the oscillatory behavior due to the presence of a third body shown in Fig.~1 of \cite{Randall:2019sab}, with the effect of a resonant transition displayed in Fig.~\ref{fig:plotInBand}.} is displayed in Fig.~\ref{fig:plotInBand}. In addition to the notable features, higher harmonics would also become more relevant as the eccentricity increases \cite{Wadekar:2023gea}. As~for the case of large tidal Love numbers \cite{Baumann:2018vus,Chia:2023tle},\footnote{A fraction of the cloud may still survive the resonant floating period(s), even for large decaying widths, which can also produce a measurable imprint in the waveforms through finite-size effects~\cite{Baumann:2018vus} (see also~\cite{Su:2021dwz}).} new dedicated templates will be needed to search for these phenomena in the GW data. 

\vskip 4pt {\bf Conclusions.} We have shown that the presence of a boson cloud surrounding a black hole in a binary system can impact the distribution of masses and eccentricities observable with GW detectors. We have also found that a greater-than-expected value of the eccentricity, $e\gtrsim 10^{-2}$ at GW frequencies $f_{\rm GW}\simeq 10^{-2}$Hz, develops for a (sub-)population of isolated stellar-mass BBHs (with ${\cal M} \lesssim 10 M_{\odot}$), right at the heart of the LISA band. Likewise, these BBHs will decay through GW emission to values of the eccentricities, i.e. $e \simeq 10^{-4}-10^{-3}$, within experimental reach of mid-band \cite{Baum:2023rwc} and Decihertz \cite{Kawamura:2020pcg} detectors. The observation of such GW signals would then provide tantalizing evidence for the existence of an ultralight particle of mass between $5\times 10^{-13}$ and $2.5 \times 10^{-12}$~eV in nature. Furthermore, we have also shown that in-band resonance transitions are possible, yielding dramatic changes in the GW frequency evolution, constituting yet another smoking-gun signature of the imprint of a boson cloud in the BBH dynamics.\vskip 4pt

There are several venues for further exploration. First, unlike Bohr-type resonances, we have concentrated here on (co-rotating) hyperfine and fine transitions which occur outside of the cloud (for the range of parameters we considered), and therefore are not subject to ionization/dynamical friction~\cite{Baumann:2021fkf,Tomaselli:2023ysb}. Preliminary studies suggest that a similar increase of eccentricity occurs for certain type of Bohr transitions at higher frequencies, which would put them within reach of the ET and CE detectors \cite{us}, but a more in-depth study is needed to take all relevant effects into account. Second, although generic, we have considered the case of uninclined  orbits. This is justified for the populations of BBHs we considered here (formed in isolation with spins parallel with the angular momentum). However, to encompass also dynamically-formed systems, we must add inclination and new (off-plane) transitions \cite{Tomaselli:2023ysb}. While our results remain unchanged for quasi-planar motion, we also expect similar conclusions to apply for inclined orbits. (In fact, as shown in \cite{Tomaselli:2024bdd, Tomaselli:2024dbw}, resonant transitions tend to equatorialize the orbit.) Finally, identical results can be drawn also for neutron star/black hole binaries. For instance, those formed in isolation have a parent black hole with mass near $M \simeq 7 M_{\odot}$, and likewise in binaries with negligible eccentricity at $f_{\rm GW} \simeq 1$Hz \cite{Belczynski:2001uc,Sedda:2020wzl,Dhurkunde:2023qoe}. The presence of a boson cloud would then also lead to larger-than-expected eccentricities, providing additional circumstantial evidence for a new ultralight particle in nature.\\

{\bf Acknowledgements.} We thank G.M. Tomaselli for informative discussions and the authors of~\cite{Tomaselli:2024bdd} for sharing a draft of their related work. The work of MB and MK is supported in part by the Deutsche Forschungsgemeinschaft (DFG, German Research Foundation) under Germany's Excellence
Strategy – EXC 2121 ``Quantum Universe" – 390833306. MB and RAP are supported in part by the ERC Consolidator Grant ``Precision Gravity: From the
LHC to LISA" provided by the European Research Council (ERC) under the European Union's H2020 research and innovation programme (grant No. 817791). 

\appendix

\section{Gravitational atom}
The range of ultralight masses that we probe yield a (very) high ``axion decay constant" ($f_a$), generically suppressing self-interactions~\cite{Yoshino:2015nsa,Gruzinov:2016hcq,Baryakhtar:2020gao,Chia:2022udn} and coupling to other species~\cite{Rosa:2017ury,Boskovic:2018lkj,Fukuda:2019ewf,Spieksma:2023vwl,Chen:2023vkq}. In addition, for the black hole masses we studied, with $5 M_\odot \lesssim M \lesssim 50 M_\odot$, we arrive at small-to-moderate values of $\alpha \lesssim 0.25$, keeping relativistic corrections to
the hydrogenic states small~\cite{Siemonsen:2022yyf,Cannizzaro:2023jle,Brito:2023pyl,Duque:2023cac}. In this regime, the eigenvalues are given by~\cite{Baumann:2018vus,Baumann:2019eav}
\begin{equation}
\begin{split}
\epsilon_{n l m}=&\,\mu\left[1-\tfrac{\alpha^2}{2n^2}-\left(\tfrac{1}{8n}+\tfrac{6}{2 l+1}-\tfrac{2}{n}\right)\tfrac{\alpha^4}{n^3}\right.\\&\left.+\tfrac{16}{2 l(2 l+1)(2 l+2)}\tfrac{\tilde{a} m \alpha^5}{n^3}\right]\,.  
\end{split}
\end{equation}
Furthermore, from Detweiler's approximation~\cite{Detweiler:1980uk,Baumann:2019eav} (with our sign convention) 
\begin{eqnarray}
-\Gamma_{nlm} \simeq 2\tilde{r}_+C_{nl}\,g_{lm} \alpha^{4l+5}(m\Omega_H-\omega_{nlm}),    
\end{eqnarray}
with $g_{lm}\equiv\prod_{k=1}^{l}\left[k^2(1-\tilde{a}^2)+(\tilde{a}m-2r_+\omega)^2\right]$, $C_{nl}\equiv\frac{2^{4l+1}(n+l)!}{n^{2l+4}(n-l-1)!}\left(\frac{l!}{(2l)!(2l+1)!}\right)^2$, $\tilde r_+= 1+\sqrt{1-\tilde a^2}$, $M \Omega_H = \tilde a/(2\tilde r_+)$;
whereas for the cloud itself decaying into GWs, we use the (non-relativistic) approximation in~\cite{Yoshino:2013ofa}. 

\section{Tidal interactions} \label{app:tidal}

For equatorial orbits and resonances triggered away from the cloud, the tidal interaction is given by~\cite{Baumann:2018vus,Baumann:2019ztm}
\begin{eqnarray} 
\bra{a} V_\star \ket{b} &\equiv& \sum^\infty_{l_\star=2} \sum_{|m_\star| \leq l_\star} \eta_{ab}^{(\star)} \E^{-i m_\star \varphi_\star} \,, \label{eq:eta_general}\\
\eta_{ab}^{(\star)} &=& - \frac{q \alpha}{r_\mathrm{c}} \mathsf{R}^{-(l_\star+1)}_\star \frac{4\pi}{2l_\star + 1}  \left|Y^\ast_{(\star)} \left(\frac{\pi}{2}, \varphi_\star \right) \right| I_r  I_\Omega \,,  \nonumber \\
I_r &\approx& \int^\infty d\mathsf{r} \mathsf{r}^2 \hat{\mathcal{R}}_b \hat{\mathcal{R}}_a \mathsf{r}^{l_\star} \, \\
 I_\Omega &\equiv& \int d\Omega Y^\ast_{a}(\theta,\phi)  Y_{(\star)} (\theta,\phi) Y_b(\theta,\phi) \,,
\end{eqnarray}
where $(\star) \equiv (l_\star,m_\star)$, $\mathsf{r} \equiv r/r_\mathrm{c}$, $\mathsf{R}_\star \equiv R_\star/r_\mathrm{c}$, $
\hat{\mathcal{R}}_c=r_\mathrm{c}^{3/2} \mathcal{R}_{c}$ is the (dimensionless) hydrogenic radial wavefunction, $Y_{lm}$ is the spherical harmonic. We leave to \cite{us} the discussion on resonances ``inside the cloud", including dipole-mediated transitions~\cite{Detweiler:2003ci,Brito:2023pyl,Duque:2023cac}. \vskip 4pt 

The Jacobi-Anger expansion applied to~\eqref{eq:eta_general} [using~\eqref{eq:small_e_orbital}],
\begin{equation} \label{eq:Jacobi_exp}
    \E^{\pm i \Delta m (\vartheta+  2 e \sin(\vartheta))}=\sum_{k=-\infty}^{\infty} (\pm 1)^k J_k(2 e \Delta m)\E^{i(k\pm \Delta m)\vartheta} \,,
\end{equation}
can be applied to the off-diagonal terms of the Hamiltonian~[cf. \eqref{eq:Schr} in the main text]. Using the properties of the Bessel function \footnote{Parity  $J_k(-x)=J_{-k}(x)=(-1)^k J_k(x)$, recurrence formulas $x\left(J_{k+1}(x)+J_{k-1}(x)\right)=2kJ_k(x)$ and the asymptotic expansion $J_{|k|}(x) \sim (x/2)^{|k|}|k|! $, $x \ll 1$.}, the tidal perturbation [cf. \eqref{eq:Schr_ecc} of the main text] becomes
\begin{eqnarray}
\eta^{(\star)}_{ab, k} &=&  \eta^{(\star)}_{ab, 0}
   \frak{f}^{\frac{2}{3}(l_\star+1)} \tfrac{(\Delta m e)^{|k|}}  {|k!|} \left(1  + \tfrac{(l_\star + 1) k }{2 \Delta m}\right) + \mathcal{O}(e^{|k|+1}) \,, \nonumber \\
   \eta^{(\star)}_{ab, 0} &=& - \frac{q \alpha}{r_\mathrm{c}} \left(\frac{r_c}{a_0}\right)^{l_\star+1}   \frac{4\pi}{2l_\star + 1} \left| Y^\ast_{(\star)} \left(\tfrac{\pi}{2}, \vartheta \right) \right| I_r  I_\Omega \,,  \label{eq:eta_ecc}
\end{eqnarray}
where $a_0=[M(1+q)/\Omega^2_0]^{1/3}$. This interaction is nonzero only if the selection rules are satisfied~\cite{Baumann:2018vus,Baumann:2019ztm}:  $-m_a+m_\star+m_b=0$, $l_b+l_\star+l_a=2p$,  $|l_a-l_b| \leq l_\star \leq l_a+l_b $. Furthermore, for equatorial orbits, for even (odd) $l_\star$ only the spherical harmonics even (odd) in $m_\star \neq 0$ are nonzero.\vskip 4pt 

\section{Atomic resonances \label{app:at_tr}}

\begin{figure*}[t!]
    \centering
      \includegraphics[width=\textwidth]{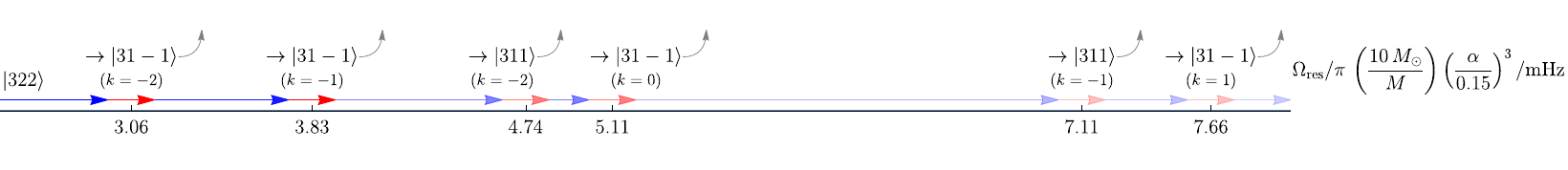}
    \caption{Orbital resonance frequencies of possible (fine) transitions for the $\ket{322}$-component of the cloud. The blue and red arrows represent vacuum and floating BBH evolution, respectively; whereas the grey arrows point to various (rapidly decaying) modes that ultimately deplete the cloud. The thinner the line the smaller the cloud gets.}
    \label{fig:plotResonances}
\end{figure*}

As the populated state has a maximal azimuthal number $m_\mathrm{max}=n-1$, at the (hyper)fine resonances, it can only transition into states with lower $m$. Such transitions are only possible on co-rotating orbits, where they obey $s\Delta m<0$, yielding floating-type motion\footnote{Notice that for the $k>|\Delta m|$ overtones, floating transitions are possible on counter-rotating orbits $(s=-1)$. The lowest one, at $k=2$, occurs for $\Delta m=-1$.}. From the selection rules for the tidal interactions, the $\ket{211}$ state has only one hyperfine transition to $\ket{21-1}$, while the only possible fine transition, to $\ket{200}$, can only occur inside the cloud. Furthermore, for small values of $\alpha$, we find that the floating time of the hyperfine transition would take longer than a Hubble time, preventing them to reach the LISA band. At the same time, (barring  a precise fine tuning of the birth frequency of the BBH+cloud) for large values of $\alpha$ we expect the $\ket{211}$ component of the cloud to decay through its own GW emission before reaching the resonant transition (see also~\cite{Takahashi:2023flk,Cao:2023fyv}). On the other hand, the (longer-lived) $\ket{322}$ state may experience various types of resonances. In contrast to early and late resonances with $k= \pm 1$, all of the $k=0$ hyperfine transitions to $\ket{32m}$ happen at the same frequency ($\Delta m$ drops out of the ratio). The dominant main (hyperfine) transition is the one to $\ket{320}$, with  $l_\star=2$ [as the $\ket{322}\to\ket{32-2}$ resonance can only be mediated by the hexadecapole ($l_\star=4$), making it extremely weak and nonadiabatic]. Transitions to the $\ket{32\pm 1}$ states are not possible for equatorial orbits. The fine resonances from the excited state are (octopolar) to the $\ket{31-1}$, $\ket{311}$ and (quadrupolar) $\ket{300}$ states, in that order in the frequency domain. We show a succession of transitions in Fig.~\ref{fig:plotResonances}. We ignored the (sinking) high-$l_\star$ Bohr resonances that can overlap with the range of frequencies that we consider here, since they do not significantly affect the dynamical evolution. We postpone the general analysis of Bohr transitions to \cite{us}.

\section{Nonlinear Landau-Zener transition} \label{app:LZ_NL_dec}

\subsubsection*{Linear solution}

The solution of the (linear) LZ transition with $\dot \Omega_k \simeq f(e) \gamma_k$ $\simeq \mathrm{const}$, including a decaying width, is given by~\cite{PhysRevA.46.4110, 1997PhRvA..55.2982V,Takahashi:2023flk} (see also~\cite{Zener:1932ws,Baumann:2019ztm})
\begin{eqnarray} \label{eq:LZ_solution_decay}
|c_a|^2 &=& \exp \left(-v_k \tau-\frac{\pi}{2} z_k \right) \left|D_{i z_k }\left(\E^{i \frac{3 \pi}{4} } \left( \tau-i v_k \right)  \right)  \right|^2 \\
|c_b|^2 &=& \exp \left(-v_k \tau-\frac{\pi}{2} z_k \right) z_k  \left|D_{i z_k -1 }\left(\E^{i\frac{3 \pi}{4}} \left( \tau-i v_k \right)  \right)  \right|^2 \,,\nonumber
\end{eqnarray}
where $D$ are
parabolic cylinder functions \cite{Zener:1932ws}, and we introduced a dimensionless time 
\begin{equation}\tau \equiv t\sqrt{|\Delta m+ k|f(e) \gamma_k }\,.\end{equation}
Notice that the physical impact of $\gamma_k$ (associated with the timescale of GW radiation) on the LZ dynamics is reflected via the ratio with the other two relevant parameters, for instance, the ratio with the (time) scale associated with the perturbation, $(\eta/\sqrt{\gamma_k})^2$, measures the adiabaticity of the transition, while $\Gamma_b/\sqrt{\gamma_k}$ describes the impact of of the decaying width relative to the dynamical time. Remarkably, ignoring backreaction effects, even though the solution changes with respect to the non-decaying case, the transition probability at infinity, given by $|c_a(\infty)|^2=\E^{-2 \pi z_k}$, turns out to be independent of $v_k$.\footnote{This fails to be the case for generic $\dot{\Omega}_k \sim t^n$~\cite{1997PhRvA..55.2982V}, namely during late inspiral.}\vskip 4pt

The presence of a decay width, however, tends to smooth the LZ transition, by transferring a fraction of the initial state earlier than the case with $v_k=0$, and also by damping late-time oscillations. Consequently, for large widths, $\tfrac{d}{d\tau}|c_a|^2$  peaks earlier than the $v_k=0$ case.  In the limit $v_k \gg z_k$, the LZ solution acquires a simple form~\cite{1997PhRvA..55.2982V,Takahashi:2023flk}
\begin{eqnarray} \label{eq:LZ_decay_approx}
|c_a|^2 &=& \exp \left[-2 z_k \left(\arctan\left(\frac{\tau}{v_k}\right)+\frac{\pi }{2}\right)\right] \,, \\
|c_b|^2 &=& |c_a|^2 \frac{z_k}{\tau^2+ v_k^2} \,.
\end{eqnarray}
In this regime, $\tfrac{d}{d\tau}|c_a|^2$  peaks at $\tau_\mathrm{max} \simeq - z_k v_k$, and the width of the transition roughly scales as 
\begin{equation}
    \frac{\Delta \Omega}{\sqrt{ |\Delta m+ k| f(e)\gamma}}  \simeq 2 v_k \sqrt{1+2 z_k^2}\,.\label{eq:tdyn}
\end{equation}
\vskip 4pt

\subsubsection*{ Balance equations}

In the two-level system with a decaying mode, we can relate the total energy and the angular momentum of the $\ket{a, b, \mathrm{BH}}$ state as [cf.~\eqref{eq:balanceE}-\eqref{eq:balance_hor_BH} in the main text]
\begin{eqnarray} \label{eq:en_ang_cloud_BH}
 (\dot L_{\mathrm{c}}+\dot S) &=& (\dot E_{\mathrm{c}}+\dot M) \frac{\Delta m}{\Delta \epsilon} \times \frac{\rho_\epsilon(t)}{\rho_m(t)}\\
\rho_x &\equiv& \frac{x_a}{\Delta x} \frac{d}{dt} |c_a|^2 - \Gamma_a |c_a|^2 - \sum \hat{\Gamma}_a |c_a|^2 \nonumber \\
&& + (a \to b) \,,
\end{eqnarray}
where $x \equiv \left\{\epsilon, m\right\}$, and the $\hat{\Gamma}$ represents other sources of dissipation, such as GW emission from the cloud~\cite{Arvanitaki:2010sy,Yoshino:2013ofa,Brito:2014wla,Arvanitaki:2014wva,Brito:2017zvb,Siemonsen:2022yyf} or ionization~\cite{Baumann:2021fkf,Baumann:2022pkl,Tomaselli:2023ysb}. From the Schr\"odinger equation [cf.~\eqref{eq:Schr}], we find
\begin{eqnarray}
\rho_x =  \frac{d}{dt}|c_a|^2 + 2|c_a|^2 \Gamma_a + 2 \frac{x_b}{\Delta x}\sum \left( \hat{\Gamma}_a |c_a|^2+\hat{\Gamma}_b |c_b|^2 \right) \nonumber \,,
\end{eqnarray}
and a similar equation applies with $a\leftrightarrow b$. After superradiance saturates the growth of the cloud in the $\ket{a}$ state, we have $\Gamma_a \simeq 0$. Moreover, for the floating time of the resonances we consider here, we can ignore other sources of dissipation during the LZ transition (setting $\hat\Gamma \simeq 0$)~\cite{us}. Hence, $\rho_x = (d/dt)|c_a|^2$, which implies $\rho_\epsilon/\rho_m=1$, yielding the expression in~\eqref{eq:balanceW}-\eqref{eq:balanceecc} of the main text.

\subsubsection*{ Nonlinear backreaction effects}

\begin{figure}
   	 \centering
   	 \includegraphics[width=0.4\textwidth]{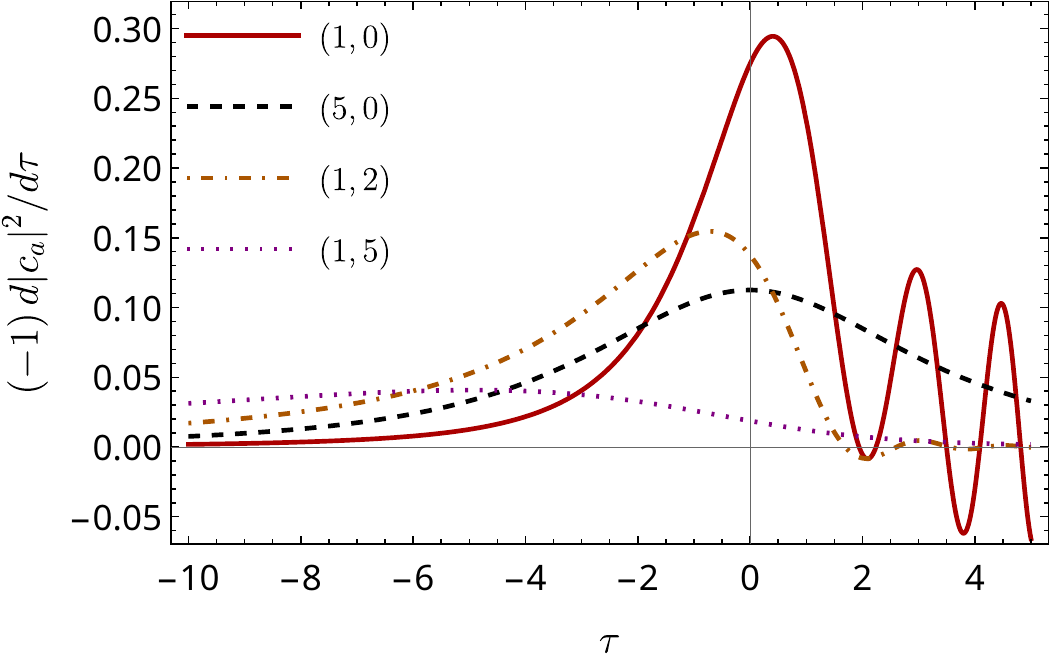}
   	 \caption{(Negative) derivative of the parent state occupancy, evaluated for a linear LZ transition. Brackets in the legend correspond to $(z_k,v_k)$. Note the decrease, and shift to the left, of the amplitude, as well as the widening of the function and damping of the late-time oscillations, as the ratio $v_k/z_k$ increases.}
   	 \label{fig_psi_LZ}
    \end{figure}

Because of the backreaction on the orbital frequency, which in turns controls the LZ transition, the problem becomes nonlinear, and it depends on the level-occupancy through the derivative of the parent state occupancy  [cf. \eqref{eq:r_of_t} in the main text]. We plot the this derivative for the linear LZ problem in Fig.~\ref{fig_psi_LZ}.
In the parts of the parameter space where it has compact support, the backreaction simply renormalizes the parameters $(z_k,v_k) \to$ $ (\zeta_k,w_k)$ [cf. \eqref{eq:balanceecc} of the main text] near the maximum value, as 
 \begin{equation} \label{eq:zeta_app} \zeta_k = \frac{\eta_k^2}{|\Delta m + k|\dot \Omega_k} = \frac{\eta_k^2}{|\Delta m + k|\gamma_k f(e) r} = \frac{z_k}{r_k(\zeta_k)}\,,
 \end{equation}
 and with the energy transfer itself depending on the level-occupancy of the cloud. The solution can, nonetheless, be found self-consistently in terms of the relevant parameters.\vskip 4pt
 
 This mapping provides a useful indicator, even in the regime of strong backreaction, to evaluate whether, e.g., floating can occur, by indicating the breakdown of the linearization of the full problem. From the self-consistency condition~\eqref{eq:zeta_app}
  the ``true'' time scale for the nonlinear problem follows $ \mathrm{T}=\sqrt{r_k } \tau $ . The LZ function derivatives can now be expressed in terms these quantities,
\begin{eqnarray}
\frac{d|c_i|^2}{d \tau} = \sqrt{\frac{z_k}{\zeta_k}} \frac{d|c_i|^2}{d \mathrm{T}}  \,,
\end{eqnarray}
leading to
\begin{eqnarray}
&& - \frac{d|c_a|^2}{d \tau} \Big|_{\tau = - \zeta_k w_k}  =   \sqrt{\frac{z_k}{\zeta_k}} \psi_k(\zeta_k,w_k) \,,\\
&& \psi_k \equiv   \frac{1}{2} \E^{-\frac{1}{2} (\pi  \zeta_k )} \Bigg[ 2 w_k \left|\mathcal{C} H_{i \zeta }\left(\frac{\mathcal{C} w_k \Delta w_k}{2} \right)\right|^2  - \nonumber \\
&& \quad  \left( \mathcal{C} H_{1-i \zeta }\left(\frac{\mathcal{C}^\ast \Delta w_k^\ast}{2} \right)  H_{i \zeta }\left(\frac{\mathcal{C} \Delta w_k}{2} \right) + \mathrm{c.c.} \right) \Bigg] \,,\nonumber
\end{eqnarray}
where $\mathcal{C}=1+i$, $ \Delta w_k = w_k (1-i\zeta_k)$ and $H$ is the Hermite function. In what follows, for the sake of notation brevity, we take the small-eccentricity approximation, with $f(e) \simeq 1$. This is also justified by the fact that the eccentricity before the LZ transition is typically small across the parameter space, which we use to evaluate the type of transition the cloud will experience. 

\subsubsection*{Negligible-decay regime} 

The $\psi_k$ function simplifies in the asymptotic regime
\begin{eqnarray}
\lim_{\zeta \to 0} \psi_k \sim   \sqrt{\pi}  \zeta \,, \,\, \And \,\, \lim_{\zeta\to\infty} \psi_k \sim   \frac{1}{4 \sqrt{\zeta}}\, \quad (w_k \ll \zeta_k) \,, \nonumber
\end{eqnarray}
in which
\begin{eqnarray}
\zeta_k \sim z_k \left(1 - \mathrm{sgn}(s\Delta m) \frac{b_k}{4 \sqrt{z_k}} \right)  \,,
\end{eqnarray}
and the value of $r_k$ given in~\eqref{rsmallv} of the main text. As advertised, the energy transfer for the adiabatic floating $(s\Delta m <0)$, is extremised by large values of $b_k$ with moderate $z_k$ parameters. In contrast, sinking $(s\Delta m >0)$ is consistent with large adiabaticity only for moderate backreaction $b_k < 4\sqrt{z}_k$. Furthermore, the large backreaction limit $(b_k \gg 1)$ of the weakly-adiabatic regime 
\begin{eqnarray} \nonumber
\zeta_k \sim \mathrm{sgn}(s\Delta m) \left(\frac{z_k }{b^2_k \pi} \right)^{1/3}  \,, \, r_k \sim \mathrm{sgn}(s\Delta m) (\pi b^2_k z^2_k)^{1/3} \,,
\end{eqnarray}
is only possible for sinking orbits. In such scenarios, the strong backreaction then mostly leads to nonadiabatic sinking transitions, with a potentially significant impact on the orbit\vskip 4pt

\subsubsection*{Strong-decay regime} 

In general, the impact of the decay width depends on the dynamical timescale of the LZ transition and should be compared with the strength of the coupling. Moreover, for eccentric orbits, even if $z_0>v_0$ at $\frak{f}_0=1$, this hierarchy may be reversed for $\frak{f}_k \neq 1$, due to the $e^{2|k|}$ suppression in $z_k$. Similarly to the case of negligible decay, we can also obtain approximate relations in the limit described by~\eqref{eq:LZ_decay_approx}, yielding
\begin{eqnarray}
\psi_k &\sim& \frac{2 \zeta_k  \E^{-\zeta_k  \left[\pi -2 \arctan(\zeta_k )\right]}}{w_k(\zeta_k^2 +1)}  \,, \\
r_k &\sim& 1 \pm 2 b_k z_k \frac{ \E^{-\zeta_k  \left[\pi -2 \arctan(\zeta_k )\right]}}{v_k(\zeta_k^2 +1)} \,.
\end{eqnarray}
 In general, this regime shares various qualitative  behavior as in the $w_k \ll \zeta_k$ case, but with the adiabaticity gain/loss and orbital impact suppressed by the ratio $b_k/v_k$.  In particular, large adiabaticity is consistent only for floating orbits, where we have
\begin{eqnarray}
\zeta_k \simeq \frac{\sqrt{2}}{\E} \sqrt{\frac{b_k z_k}{v_k}} \,,\quad r_k =  \frac{z_k}{\zeta_k} \to 0 \,. \quad (b_k \gg 1) \nonumber
\end{eqnarray}
In contrast, for sinking orbits, the strong backreaction requires a small population transfer, and we find
\begin{eqnarray}
\zeta_k \simeq \frac{v_k}{2b_k}\,,\quad r_k \simeq \frac{2 b_k z_k}{v_k} \gg 1\,. \quad (b_k \gg 1) \nonumber
\end{eqnarray}

Notice that, somewhat counter-intuitively, a strong-decay width not only does not necessarily imply the total depletion of the cloud, instead it suppresses~$r_k$, hence the ability of the system to float, resulting in a {\it lesser} amount of the cloud being transferred to the decaying mode (see \eqref{eq:cloud_leftover} below).

\section{Floating}\label{app:ecc}

For values of the energy transfer $r_k \lesssim 0.2$, the growth of the orbital frequency is sufficiently suppressed to allow for the possibility of floating. In that case, the growth of (initially small) eccentricity is given, in units of the dynamical time introduced in \eqref{eq:tdyn}, by
\begin{eqnarray}
e(t) &\simeq& \sqrt{e^2_\mathrm{in} + I_e \tau (1- \frak{f}_k)} \,, \quad \quad \quad (\frak{f}_k \neq 1) \label{eq:ecc_of_t_overtone} \\
e(t) &\simeq& e_\mathrm{in} \exp{-\frac{11}{6} I_e \tau} \,, \quad \quad \quad\,\, (\frak{f}_k = 1)  \\   I_e &\equiv&  \frac{2}{3}\frac{\sqrt{\gamma_0}}{\Omega_0}  \frac{ \frak{f}^{5/6}_k }{|\Delta m + k|^{1/2}} \,. \nonumber
\end{eqnarray}
The critical points of the evolution of the eccentricity depend both on $|\Delta m|$ and $|k|$. For the first few values of $|k|$ they are described by the polynomial fit $e_\mathrm{cr}=g_0+g_1 k +g_2 k^2$, where $g_0=\{0.3,0.2,0.16\}$, $g_1=\{0.18,0.16,0.14\}$, $g_2=\{2,1.6,1.3\}\times 10^{-2}$, all for $|\Delta m|=\{1,2,3\}$ in respective order.\vskip 4pt

\begin{figure*}
   \begin{tabular}{cc}
\includegraphics[width=.37\textwidth]{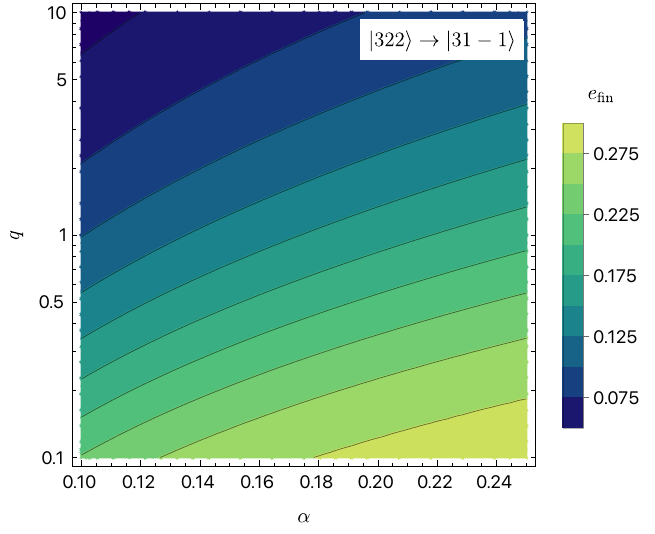}
\qquad
\includegraphics[width=.4\textwidth]{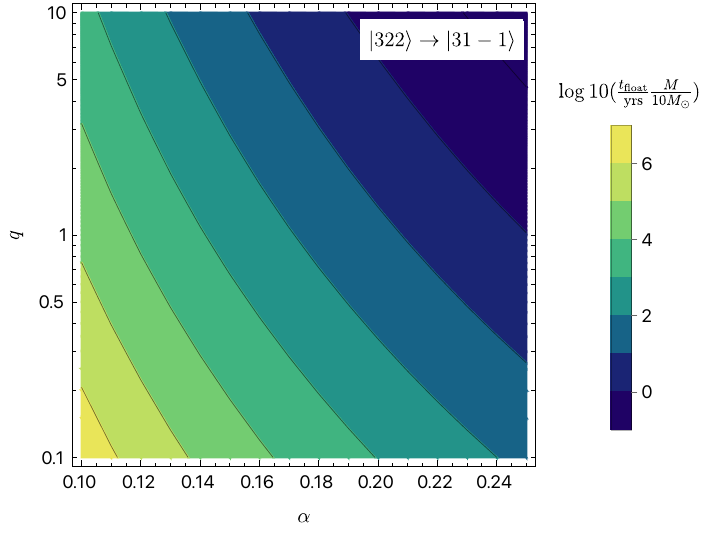}
\end{tabular}
\caption{Final eccentricity ({\it left}) and floating time ({\it right}), for the $\ket{322}\to\ket{31-1}$ transition with $k=-1$, assuming initial conditions that lead to a (long-lasting) floating orbit ($r_{-1}\simeq 0$) and $e_\mathrm{in} \ll 1$.}
	\label{fig:plotsefloat}
\end{figure*}

To calculate the evolution of the eccentricity towards the fixed point, we change the time variable in the evolution equations [cf. \eqref{eq:balanceecc}], from $t$ to $e^2(t)$, yielding
\begin{eqnarray} \label{eq:ecc_integral}
\int^{e^2_\mathrm{fin}}_{e^2_{\mathrm{in}}}  \frac{d(e^2)}{  \sqrt{1-e^2} \left[\frac{g(e)}{f(e)} -\frak{f}_k\right]} = I_e b_k\left(1 - |c_a(\infty)|^2  \right) \,,
\end{eqnarray}
The result can then be approximated by~\eqref{growth-ecc} in the main text, where
\begin{eqnarray}
\mathrm{C}_k=c_{\Delta m, k} \left[ I_e b_k \left(1 - |c_a(\infty)|^2 \right)  + \frac{ e^2_\mathrm{in}}{1- \frak{f}_k}  \right]  \,,
\end{eqnarray}
and $c_{\Delta m, -1}=$ $ \{2.37, 2, 1.37\}$ for $\Delta m = \{-1,-2,-3\}$. From the value of $e^2_\mathrm{fin}$ in~\eqref{eq:ecc_integral}, we can also estimate the duration of the floating period
\begin{eqnarray}
 \Delta \tau_\mathrm{Fl} &=& \frac{1}{I_e} \int^{e^2_\mathrm{fin}}_{e^2_{\mathrm{in}}} d(e^2) \frac{1}{\sqrt{f(e)}  \sqrt{1-e^2} \left[\frac{g(e)}{f(e)} -\frak{f}_k\right]} \,, \nonumber\\
 &\simeq& \frac{\mathrm{C}_k}{\Tilde{c}_{\Delta m, k}I_e  }  \,, \quad \frac{c_{\Delta m, k}}{\Tilde{c}_{\Delta m, k}}=\frac{1}{\sqrt{f(e_\mathrm{cr})}}\,.
 \label{eq:tfloat}
\end{eqnarray}
An exemplary parameter space of final eccentricity and floating time is shown in Fig.~\ref{fig:plotsefloat}.

\vskip 4pt
Strong floating provides a distinct phase of the non-linear LZ transition, during which typically most of the population transfer occurs. From~\eqref{eq:r_of_t} of the main text, we have
\begin{eqnarray}
\frac{d|c_a|^2}{d\tau} = \frac{r(\tau)-1}{b_k} \,,
\end{eqnarray}
yielding a linear-in-time decay of the population during floating, and a transfer of population given by
\begin{equation}
|c_a(\infty)|^2  \simeq |c_a(\tau <\mathrm{Fl})|^2 - (1-r_k)\frac{\Delta \tau_{\rm Fl}}{b_k} \,.   \label{eq:ca}
\end{equation}
%

In the strong decay regime, the condition $r(t) \to 0$ at the resonance is necessary at each point in time, but not sufficient to guarantee a steady floating-type period. In addition, there must be enough of the cloud left to sustain a small $r(t)$. Following~\cite{Takahashi:2023flk}, one can estimate the sufficient condition by considering the minimum amount of cloud needed to ``startjump'' a floating period. Applying the linear LZ solution~\eqref{eq:LZ_decay_approx} in ~\eqref{eq:r_of_t} of the main text, we find
\begin{eqnarray} \label{eq:cloud_min}
|c_a|^2_\mathrm{min} \simeq \frac{v_k\sqrt{f(e_\mathrm{in})}}{2 b_k z_k(e_{\rm in})} (1-r_k) \,.
\end{eqnarray}
The left hand side can be interpreted as the minimal amount of cloud needed to start floating at a particular resonance. In turn, if the right hand side is larger than one, floating cannot start. The same condition can also be used to estimate the amount of cloud left when floating stops, by matching into the linear LZ solution {\it backwards}, from the end of the floating time,
\begin{eqnarray} \label{eq:cloud_leftover}
|c_a(\infty)|^2 \simeq \frac{v_k\sqrt{f(e_\mathrm{fin})}}{2 b_k z_k(e_\mathrm{fin})} (1-r_k) \,,
\end{eqnarray}
which then becomes the portion of the cloud surviving {\it after} floating stops. Strong decay and small $z_k$ could in principle interrupt the floating period and leave a moderate amount of the cloud intact.\footnote{This is consistent with the ``resonance breaking" phenomena discussed in~\cite{Tomaselli:2024bdd}. We thank the authors of~\cite{Tomaselli:2024bdd} for bringing it to our attention.} However, at the overtones we have $z_k \sim e^{|2k|}$, which is increasing during floating. Hence, as the eccentricity approaches the critical point, for instance at the $k=-1$ overtone, the value of $z_k$ increases by a factor $\left(e_{\rm cr}/e_{\rm in}\right)^2 \simeq 10^{2}-10^{3}$, thus significantly extending the floating period, and reducing the amount of cloud left after the transition, in comparison with the na\"ive estimate in~\eqref{eq:cloud_min}.\vskip 4pt 

In general, the equations in~\eqref{eq:ecc_integral}, \eqref{eq:tfloat} and~\eqref{eq:cloud_leftover} must be solved self-consistently in order to determine the end state of floating. As an estimate, we may apply~\eqref{eq:ecc_of_t_overtone} to~\eqref{eq:cloud_leftover}, and assuming $e_\mathrm{in} \ll 1$,
we have
\begin{eqnarray}
&& \frac{\Delta \tau_\mathrm{Fl}}{b_k/\sqrt{f(e_\mathrm{cr})}} \simeq  \left(\frac{x-1}{2x} +\sqrt{\frac{(x+1)^2}{4x^2}-\frac{\lambda}{x}} \right) \, ,\\
&& \lambda = \frac{v_k}{2 b_k z_k(e_\mathrm{in})}  \,, \quad x = \frac{I_e b_k (1-f_k)}{e^2_\mathrm{in}}\nn \,.
\end{eqnarray}
Notice that for the $k=-1$ overtone, the dependence on $e_\mathrm{in}$ drops out from the ratio $\lambda/x$. In this case $\lambda \ll x$, and we find the longest periods of floating, $\Delta \tau_{\rm Fl} \simeq b_k/\sqrt{f(e_{\rm cr})}$, and largest depletion of the $|c_a(\infty)|^2 \lesssim r_k$. Depending on the resonances and the parameter space, such hierarchy may be also valid for higher overtones.

\begin{figure*}[t!]
\begin{tabular}{cc}
\includegraphics[width=.4\textwidth]{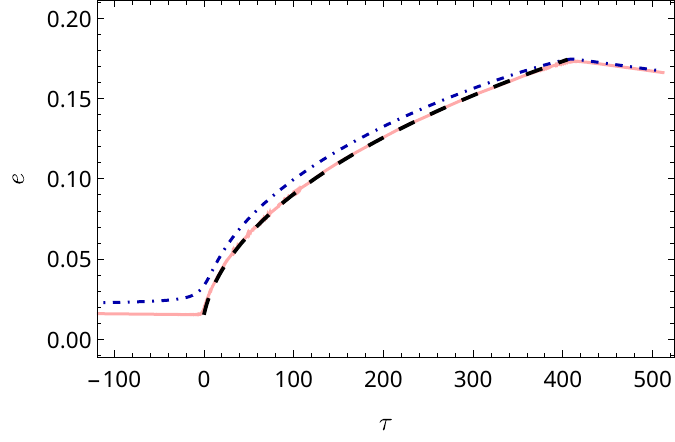}
\qquad
\includegraphics[width=.4\textwidth]{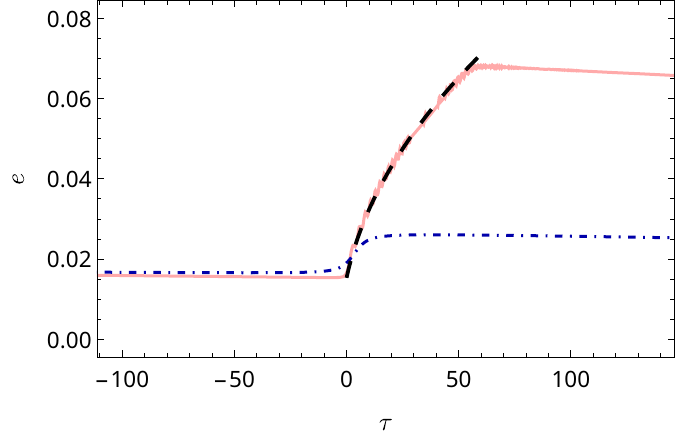}
\end{tabular}
\caption{Numerical evolution of the examples described in App.~\ref{app:num}. The figures correspond to the same $z_k$ but different values for $b_k$, with the left plot having a stronger backreaction parameter (a few times larger) than on the right. The curves show the eccentricity evolution for vanishing (pink) and large decay $v_k \gg z_k$ (blue, dot-dashed), both yielding floating orbits ($r_k \simeq 0$), except for the case of a large decay with small backreaction (blue curve on the right panel has $r_k \simeq 0.8$). The semi-analytic solution (black, dashed) [cf.~\eqref{eq:balanceecc} in the main text], is in remarkable agreement within its regime of validity ($r_k \simeq 0$).}

	\label{fig_e_num}
\end{figure*}

\section{  Numerical validation} \label{app:num}

Our semi-analytic approach,  described in Apps. D and E, rely in part on the small initial eccentricity of the BBHs prior to entering the transition region. We have validated this by numerically solving the Schr\"{o}dinger equation for arbitrary (planar) orbits, coupled with the energy-momentum balance equations [cf.~\eqref{eq:Schr},\eqref{eq:balanceE}-\eqref{eq:balance_hor_BH} of the main text], for a number of representative examples, including a broad inital eccentricity range $10^{-2} \lesssim e_\mathrm{in} \lesssim 0.5$. For this purpose, we used the generic description of the orbital evolution, $R_\star = a(1-e \cos{u})$, as a function of the eccentric anomaly $u$, which is related to the mean anomaly via Kepler's equation $u-e\sin(u)=\vartheta$ (e.g.~\cite{Tremaine_Dynamics}). We used the NDSolve routine in Mathematica~\cite{Mathematica}, monitoring the  violation of unitarity in the $\Gamma_b=0$ regime, residual for $\Gamma_b \geq 0$, as well as the state occupation numbers~\eqref{eq:LZ_solution_decay}. We have also checked the consistency of our results by comparing with the values in~\cite{Baumann:2019eav} for $(e,\Gamma_b)=0$. \vskip 4pt

In general, the numerical results are broadly consistent with the analytical arguments. As expected, the largest discrepancy occurs for the estimates of $r(t)$ and $|c_a(\infty)|^2$. For the floating regime, the analytic results tend to overestimate their respective values by a factor of few to an order of magnitude. Hence, our conclusions based on the analytic approximations {\it err on the side of caution}. In Fig.~ \ref{fig_e_num} we present (left) the numerical eccentricity evolution of a two-level system with $\Delta m=2$ for  values $\{z_0=4, v_0=2, b_0 =200, e_0 = 10^{-2}\}$, at the main resonance. 
For these parameters, we have $\{z_{-1} = 7 \cdot 10^{-3},v_{-1}=3.4, b_{-1}=450, I_e = 2 \cdot 10^{-4}\}$ at the $k=-1$ overtone, for which one finds a strong-decay regime: $v_{-1} /z_{-1}= 512$. We start the numerical evaluation before the first overtone, following also the case $v_{-1}=0$. Away from the resonance, the orbital evolution closely follows standard GW evolution in vacuum. The approximations from Sec.~\ref{app:LZ_NL_dec} then correctly predict the strong floating that occurs (in both setups) at the frequency given by~\eqref{eq:freq_overtone}. Notice that, modulo small deviations, both cases follow the $r_k \simeq 0$ prediction for the eccentricity growth in~\eqref{eq:ecc_integral}.  Furthermore, broadly consistent with our estimates, the left-over occupancy of the inital state is given by $|c_a(\infty)|^2 \simeq 10^{-4}$ and $|c_a(\infty)|^2 \simeq 10^{-3}$, with and without the decay, respectively. Finally, reducing the backreaction to $b_0=25$ (Fig.~\ref{fig_e_num}, right), the decaying case does not develop floating, as $r_k \simeq 0.8$, although as expected the eccentricity will still grow slightly and the transfer of population is somewhat increased compared to the linear LZ transition. In contrast, the non-decaying case still exhibits a strong floating, follows the predicted growth of the eccentricity, and transfers the parent state up to $|c_a(\infty)|^2 \simeq 5 \times 10^{-3}$.

\begin{figure*}[t!]
\begin{tabular}{cc}
\includegraphics[width=.4\textwidth]{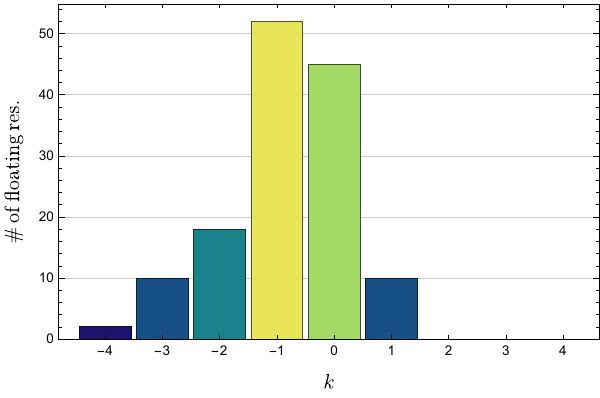}
\qquad
\includegraphics[width=.4\textwidth]{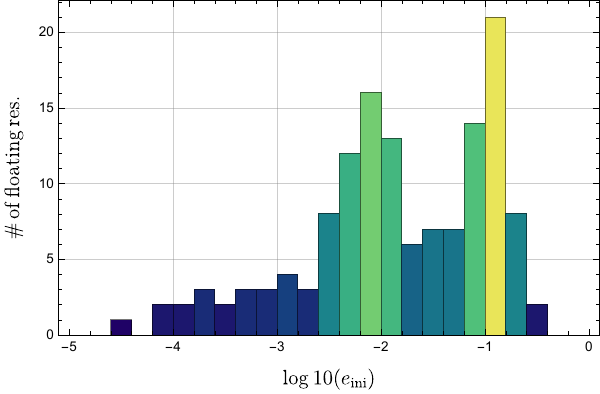}
\end{tabular}
\caption{Number of floating resonances for specific $k$ (\textit{left}) and binned by initial eccentricity (\textit{right}) for the floating orbits the population shown in Fig.~\ref{fig:Populations} experiences.}
	\label{fig_k_val_e_ini}
\end{figure*}

\begin{figure}
    \centering
    \includegraphics[width=0.4\textwidth]{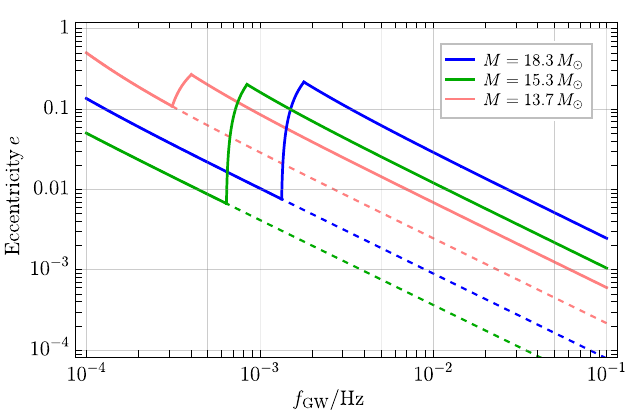}
    \caption{Eccentricity evolution (in terms of the GW frequency) for representative examples from Fig.~\ref{fig:Populations} in the main text. Solid lines show the evolution with the cloud, while dashed lines represents the standard vacuum evolution.}
    \label{fig:plotEcc}
\end{figure}

\begin{figure*}[t!]
\begin{tabular}{cc}
\includegraphics[width=.41\textwidth]{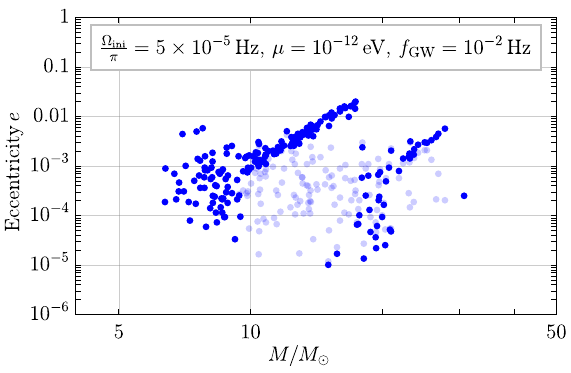}
\qquad
\includegraphics[width=.39\textwidth]{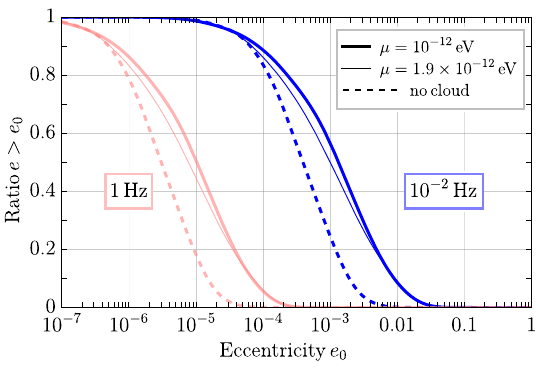}
\end{tabular}
    \caption{Same as Fig.~\ref{fig:Populations} in the main letter, but with $\Omega_\mathrm{ini}/\pi=5\times10^{-5}\,\mathrm{Hz}$ as the birth orbital frequency for the BBH+cloud system.}
    \label{fig:Populations5em5}
\end{figure*}

\section{Eccentricity distribution\label{app:parameter}}

To evaluate the distribution of eccentricities in the population from \cite{Breivik:2016ddj}, we order (in frequency) all possible hyperfine and fine resonances with the corresponding overtones, up to $|k| = 5$. (A subset is displayed in Fig.~\ref{fig:plotResonances}.) We then calculate the corresponding parameters $(z_k, v_k, b_k)$ for every resonance the cloud may encounter. The condition that the initial state had enough time to reach the resonance is  imposed, constrained by the lifetime imposed by its own GW emission. We evolve in vacuum via \cite{Peters:1963ux,Peters:1964zz} between transitions. At every resonance, we  estimate $r_k$ as explained in App.~\ref{app:LZ_NL_dec}, and from there the floating time, eccentricity growth, and left-over cloud, as explained in App.~\ref{app:ecc}. We use a cutoff at  $10^{-3}$ of the initial $M_{c,0}/M$, to estimate when the cloud's influence on the orbital dynamics becomes negligible.\vskip 4pt

For the BBHs shown in Fig.~1 of the main letter, only a few percent experience a floating transition $\ket{322} \to \ket{31-1}$  with $k<-1$, while $25\,\%$ float at the $k=-1$ resonance, making it the dominant one. Approximately $15\,\%$ of the population float at the $k=0$ transition. Only a few experience resonances to $\ket{311}$ or $\ket{300}$ states. The black holes with the largest masses, hence largest values of $\alpha$ for fixed $\mu$, can float at the $\ket{320}$ hyperfine transition, where of the order of $5\,\%$ (each),  see an earlier resonance and increase eccentricity there; or see the main or $k=1$ resonance and decrease eccentricity instead. We do not find a significant impact for overtones with $|k|>4$. We show in the left panel of Fig.~\ref{fig_k_val_e_ini} the distribution of different $k$'s experienced by the BBH population we studied in the main text. We show in the right panel the distribution of initial eccentricities before floating starts. As a consistency check, notice that for most of the resonances we have $e_\mathrm{in} \lesssim 0.1$, which further support the validity of the small-eccentricity approximation. In Fig.~\ref{fig:plotEcc} we plot the eccentricity evolution for three representative cases drawn from the population in Fig.~\ref{fig:Populations} of the main text.

 \begin{figure}
   	 \centering
   	 \includegraphics[width=0.4\textwidth]{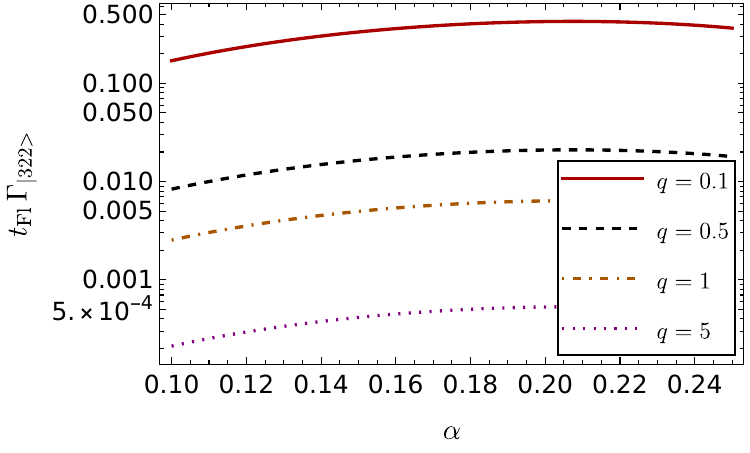}
   	 \caption{Ratio of floating and (excited-state) decay time for the $k=-1$ overtone of the fine transition, assuming the parent black hole's spin is reduced to  $\tilde{a}=0.9 \tilde a_s$, with $\tilde a_s = 2\alpha$ the value at saturation.}
   	 \label{fig_flt_dec_time}
    \end{figure}

\subsubsection*{Birth frequency}

Similar to Fig.~\ref{fig:Populations} in the main letter, we show  in Fig.~\ref{fig:Populations5em5} the eccentricity distribution at $ f_{\rm GW} = 10^{-2}\,$Hz, but with a lower birth frequency for the BBH+cloud system, at $\Omega_\mathrm{ini}/\pi=5\times10^{-5}\,$Hz. Although the number of larger-than-expected eccentricity points is somewhat lower, the plot illustrates the robustness of our predictions against changes in the initial conditions. In general, a large portion of BBHs that merge within a Hubble time will be born with orbital frequencies between $[0.5,1] \times10^{-4}$Hz ~\cite{Kowalska_2011}. Moreover, BBHs that are formed through a common-envelope mechanism that can bring the binary closer, e.g. \cite{Belczynski:2017gds}, at the same time may produce a (younger) secondary black hole that can ultimately carry the boson cloud, pushing the cloud's birth frequency towards higher values, thus avoiding altogether ($k=0$) hyperfine transitions that can decrease the eccentricity. Overall, our findings demonstrate that a skewed-type distribution and larger-than-expected eccentricities are robust predictions of the presence of boson clouds in the BBH dynamics.  

\section{Background (black hole) parameters}

Finally, let us comment on the backreaction on the black hole mass and spin.
We will implement this in detail in a forthcoming paper. However, it is worth emphasising already that this effect does not change the main message of the letter: fixed points in the eccentricity evolution and the imprint in its final distribution at observable GW frequencies. We provide here a brief argument to support this claim. \vskip 4pt First of all, the changes in $\alpha$ (through the growth of the black hole mass) move the position of the resonances towards higher frequencies, which may help in sustaining the floating condition. Regarding the changes in spin, while they depend on the sign of $m_b$, most of transitions tend to decrease the value of the parent black hole's spin, which in turn affects the superradiant condition. As a result, the $\ket{322}$ state may develop a decay width. However, even in the most extreme scenario, and assuming the entire cloud is instantly reabsorbed, we would still have $\Gamma_{\ket{322}} \simeq \alpha^{13}$ (with $\ell=2$ for the excited state). As a consequence, as we show in Fig.~\ref{fig_flt_dec_time}, the decay time turns out to be longer than the floating time for all of the fine transitions. This effectively allows us to ignore the resulting width of the $\ket{322}$ state in the determination of the growth of eccentricity.\vskip 4pt 

For the hyperfine transitions, on the other hand, the situation is more subtle, since the floating times are in principle longer. In that case, one has to take into account that the decrease of the spin does not occur instantaneously, but on the time-scales of the LZ transition governing the population of the $\ket{b}$ state. 
Preliminary studies indicate that the qualitative results for the hyperfine resonance are not significantly modified. In particular, since hyperfine transitions fall typically in the weak-decay regime, the floating condition is expected to be generally robust.  This expectation is also in agreement with the results of~\cite{Takahashi:2023flk}, where the hyperfine transition $\ket{211} \to \ket{21-1}$ was considered including the evolution of the background parameters. The authors did not find a significant disruption of the floating mechanism due to the absorption of mass and angular momentum. Notice, in addition, that for the excited state we have $\Gamma_{\ket{322}} \ll \Gamma_{\ket{211}}$, which further supports this conclusion. We postpone a full analysis to \cite{us}.

\newpage
\bibliography{references.bib}

\end{document}